\documentclass[twocolumn,twoside,slac]{revtex4}
\ifx\pdftexversion\undefined
  \usepackage[dvips]{graphicx}
\else
  \usepackage[pdftex]{graphicx}
\fi

\usepackage{fancyhdr}
\usepackage{xspace}

\begin{document}

\newcommand{\cwASK}{\textsc{ask}\xspace}
\newcommand{\cwASKCap}{\textsc{Ask}\xspace}
\newcommand{\cwAthena}{\textsc{athena}\xspace}
\newcommand{\cwAthenaCap}{\textsc{Athena}\xspace}
\newcommand{\cwAtlas}{Atlas\xspace}
\newcommand{\cwAtlasCap}{Atlas\xspace}
\newcommand{\cwAtlasFull}{A Toroidal LHC ApparatuS\xspace}
\newcommand{\cwCVS}{\textsc{cvs}\xspace}
\newcommand{\cwCVSCap}{\textsc{Cvs}\xspace}
\newcommand{\cwGanga}{\textsc{ganga}\xspace}
\newcommand{\cwGangaCap}{\textsc{Ganga}\xspace}
\newcommand{\cwKerberos}{\textsc{kerberos}\xspace}
\newcommand{\cwKerberosCap}{\textsc{Kerberos}\xspace}
\newcommand{\cwPython}{\textsc{python}\xspace}
\newcommand{\cwPythonCap}{\textsc{Python}\xspace}
\newcommand{\cwUnix}{\textsc{unix}\xspace}
\newcommand{\cwUnixCap}{\textsc{Unix}\xspace}

\newcommand{\cwRefFigure}{Fig.$\!$}
\newcommand{\cwRefFigureCap}{Figure}
\newcommand{\cwRefSection}{Section}
\newcommand{\cwRefSectionCap}{Section}

\newcommand{\cwcf}{c.f.\xspace}
\newcommand{\cweg}{e.g.\xspace}
\newcommand{\cwetc}{etc.\xspace}
\newcommand{\cwie}{i.e.\xspace}
\newcommand{\cwndf}{n.d.f.\xspace}
\newcommand{\cwrms}{r.m.s.\xspace}
\newcommand{\cwvs}{v.s.\xspace}
\newcommand{\cwwrt}{w.r.t.\xspace}

\title{The Athena Startup Kit}

%

\author{W.~T.~L.~P.~Lavrijsen}
\affiliation{LBNL, Berkeley, CA 94720, USA}

\begin{abstract}
The Athena Startup Kit (\cwASK), is an interactive front-end to the
\cwAtlas software framework (\cwAthena).
Written in \cwPython, a very effective ``glue'' language, it is build on top
of the, in principle unrelated, code repository, build, configuration, debug,
binding, and analysis tools.
\cwASKCap automates many error-prone tasks that are otherwise left to the
end-user, thereby pre-empting a whole category of potential problems.
Through the existing tools, which \cwASK will setup for the user if and as
needed, it locates available resources, maintains job coherency, manages the
run-time environment, allows for interactivity and debugging, and provides
standalone execution scripts.
An end-user who wants to run her own analysis algorithms within the standard
environment can let \cwASK generate the appropriate skeleton package, the
needed dependencies and run-time, as well as a default job options script.
For new and casual users, \cwASK comes with a graphical user interface; for
advanced users, \cwASK has a scriptable command line interface.
Both are built on top of the same set of libraries.
\cwASKCap does not need to be, and isn't, experiment neutral.
Thus it has built-in workarounds for known gotcha's, that would otherwise be a
major time-sink for each and every new user.
\cwASKCap minimizes the overhead for those physicists in \cwAtlas who just want
to write and run their analysis code.
\end{abstract}

\maketitle

\thispagestyle{fancy}


\section{\label{sec:intro}INTRODUCTION}

\cwAtlasCap~\cite{ref:Atlas94} is a multi-purpose high energy physics
experiment planned for the Large Hadron Collider~\cite{ref:LHC95}~(LHC), which
is scheduled for startup after 2007.
The data processing needs of \cwAtlas, its variety in subsystems, and its
expected longevity pose a big challenge for the software that will be needed
for extracting physics out of the future \cwAtlas data.
This challenge has attracted a large software community, with developers
working today, still a few year before the experiment starts running, on every
aspect of the system.
Most of the huge \cwAtlas code base, from the software that the end-user runs
to the configuration of the infrastructure tools, is currently under heavy
development.
This has resulted in a state of flux, as a consequence of which all software
usage becomes low-level, and all users end up fixing the same problems
multiple times.
The threshold for new users is high, and casual users are, time and again,
forced to relearn most of what they thought they knew.

A large fraction of the \cwAtlas software runs within the \cwAthena framework,
and in order to alleviate several of the chores that developers and end-users
alike have to perform in order to work with that system, the Athena Startup
Kit (\cwASK) has been developed.
Its main features are:

\begin{itemize}

\item
{\it Automation} of end-user tasks, to save time.

\item
{\it Integration} of the framework, the releases, and the release tools to
improve perceived coherency.

\item
{\it Encapsulation} of volatile wisdom (from web-pages and peoples' heads) of
known problems and their workarounds, to make it more accessible.

\end{itemize}

There are several different categories of \cwAthena users and \cwASK addresses
the needs of all of them.
\cwASKCap performs the steps needed to setup tools, locate resources, work
around known problems, and execute the \cwAthena program.
In addition, \cwASK has \cwAthena specific functionality: it can generate and
update user code meant for \cwAthena execution (``algorithms''), generate
default job options scripts, and update the dependencies for different releases
and compile modes.

A simplistic implementation of \cwASK would mean that its maintenance would
turn into a full-time job, which is unacceptable.
Thus, \cwASK is coded in such a way that all expected changes are properly
taken into account when they occur, that a solution that has been
successfully applied before is applied again if the problem resurfaces, and
that it has a ``Plan B'' for most tasks.
This paper describes in detail how this robustness is achieved.

\section{WORK MODEL}

There are three main aspects to everything that \cwASK performs: first, setup
a clean-room to work in; then, decompose the task at hand; and finally, always
be ready to handle failure.

\subsection{The Clean-Room}

There is a whole score of problems related to improper configuration of the
working environment by the end-user.
In order to keep full control of the environment, \cwASK does not require any
actions outside its own process (or any of its subprocesses).
Although the user needs to have full access to the environment in case all else
fails, in general she is shielded.
\cwASKCap accomplishes this by executing a working shell in the background, to
which it establishes open pipes for input and output, and it then works on this
shell as if it was itself just another end-user.
This shell is its ``clean-room,'' and \cwASK has full control: all
communication proceeds through an API, which allows \cwASK to verify and
validate any commands, their progress, and their results.
It is not uncommon to allow commands that may cause havoc, and simply correct
settings afterwards.

A shell that is started from the current process inherits the environment and
configuration of the current process, which can be invalid to begin with.
Note however, that what may seem like an invalid configuration to \cwASK may in
fact be an experimental setup from an expert user, who really wants to have the
environment setup this way in order to see what happens.
Therefore, \cwASK can not simply override any existing configuration.
Instead, it accepts the settings from the user and only provides what appears
to be missing.
This is a very productive and pleasant way of working: as an expert user, you
configure only that part of the system that you are interested in, and \cwASK
takes care of the rest.

In practice, it turns out that for beginning and casual users a different
approach is needed: many of them have residual configuration in their login
files.
This usually does not cause any problems at first, but sooner or later these
settings will go out of date, leaving the user puzzled when things start to
mysteriously fail.
The user will, typically, blame\footnote{In fact, it used to be that {\em
every} \cwASK bug report was resolved by simply having the user clean up his
environment.} this on \cwASK and may decide to drop the tool in
disappointment.
Consequently, \cwASK comes in two modes: expert and non-expert (the latter is
the default).
In non-expert mode, \cwASK removes any known \cwAtlas software configuration
from the environment before executing any of its own code.
This way, \cwASK will operate as if it started from a clean environment and
perform the whole setup and configuration itself.

Regardless of whether the user or \cwASK has done the configuration, every task
is executed from a known state: \cwASK enters a base directory, from where the
task can move into its working directory, and it locates the required
resources and verifies their configuration, repairing the setup as needed
(taking guidelines from experts, overriding non-experts).
Each task can then be implemented in a straightforward way, since it will, by
design, not get executed until it has a very good chance of succeeding.

If an environment setup is irreparably broken, or if a critical resource is
unlocatable, \cwASK has no choice but to stop.
However, since such a problem occurs in very specific cases, it is often (but
not always) possible to present a clear and concise error report to the user.
See \cwRefSection~\ref{sec:error} for a more detailed discussion of error
handling by \cwASK.

\subsection{Task Decomposition}

Many of the tasks performed by \cwASK use the same tools and require similar
resources.
Further, steps within individual tasks sometimes overlap and can consequently
be shared.
In fact, most tasks share the same initialization and finalization step.
Thus, it makes sense to implement the tasks in a layered execution model:

\begin{itemize}

\item {\bf Encapsulate tools}\\
Each task has access to the execution shell, but only standard \cwUnix commands
are executed directly.
Tools that are part of \cwAtlas software, however, are normally\footnote{Similar
to the setup of the environment, the user can access the tools directly if all
else fails} used directly.
Instead, tools are encapsulated in modules, which enables \cwASK to locate tools
and to verify their setup and possibly update or repair it, before any attempt
is made to perform actions with the tool.
This is the lowest layer.

\item {\bf Refactorize steps}\\
It is a good programming practice to minimize the amount of duplicate code and
in \cwASK this means that all steps that are shared between tasks are
refactored.
Further, the order of execution of steps is also often the same.
By design many actions need the same set of parameters, and it is therefore
straightforward to use meta-programming to create templates that are
parametrized on the core action of a task.
This guarantees that the execution order is proper, which greatly simplifies the
implementation of individual steps.
This is the programmatic layer.

\item {\bf Construct recipes}\\
The end-user typically does not think about tasks in the same granularity as
tool authors do.
Thus, recipes, that are end-user tasks, are constructed from the programmatic
tasks.
Since programmatic tasks are supposed to be self-contained, they can be freely
mixed and matched in the recipes: it is only the goal of the recipes that puts
constraints on their order.
This is the user layer.

\item {\bf User interfaces}\\
The recipes typically take input and have output.
The former needs to be collected from the user, and the latter needs to be
presented back to the user.
The user interfaces connect input/output elements with recipes to present the
\cwASK functionality to the user.
There is a graphical interface and a command line interface.
This is the application layer.

\end{itemize}

Third party programs or scripts, can access \cwASK functionality at any layer,
but do so most commonly at the programmatic and/or user layer.
These programs are typically similar in setup as the user interfaces (in effect,
they are also located in the application layer).
For example, the \cwASK distribution includes a program that performs full
\cwAtlas software builds and is targeted at software librarians.
But in principle, it is just a batch version of the command line interface with
a few different defaults and a few extended recipes.

\subsection{\label{sec:error}Error Handling}

There are several different kinds of errors that \cwASK can encounter:

\begin{itemize}

\item{\bf Random.}\\
Errors due to (temporarily) missing resources such as disk access,
\cwKerberos tokens, the tag collector, etc.
These errors come and go and, unfortunately, the solution is usually
``come back later, and try again'' or requires some user intervention.
Reporting is often, but not always, straightforward.
For example, if a \cwKerberos ticket is missing and \cwASK attempts to access
a file in the code repository, it will be \cwCVS that fails and \cwASK will
only have an error from that tool to report, thus masking the real error.

\item{\bf User actions.}\\
Errors because of a file that does not compile, an executable that does not
link, an option script that the user requested but that does not exist, etc.
Reporting these errors is simple: \cwASK needs to retrieve the error code,
stop the task, and report the error message from the tool that failed.

\item{\bf Broken system.}\\
Errors as a result of misconfiguration of the \cwAtlas software distribution
that is used.
There is not much point in reporting this kind of error, unless the user is
testing the distribution.
Instead, \cwASK attempts to work around the problem.
If that fails, then the requested task can not be performed.

\end{itemize}

It is the last item that is the most interesting, because in handling that
category of errors there is the most to gain in terms of productivity for the
end-user, especially if he or she is a novice or casual user.
In order to deal with broken systems, \cwASK has several options.
First of all, it may come prepared: for several releases and configurations,
it comes with specific workarounds that are simply programmed into \cwASK.
That way, the problem will not appear in the first place, because the
workaround is the default.

It gets more interesting when a task does fail.
In that case, \cwASK will try a few more things that might work, such as
locating an alternative resource, querying an alternative source, or simply
accepting a default and hope for the best.
The ticket here is that this is precisely how an end-user would attempt to
tackle the problem herself, with the exception that \cwASK can perform the
trials much, much faster, and it has therefore a good chance of arriving at
a solution with the end-user never knowing that there was a problem.

As a final resort, \cwASK sometimes explicitly fixes a workspace.
Since this usually involves moving or replacing files and/or directories, this
is done only when the case is very clear.

Note that as before, when all else fails, \cwASK does allow expert access at all
levels, at any time.

\section{IMPLEMENTATION}

One of the stated goals of \cwASK (see section~\ref{sec:intro}) is the
integration of the framework, the releases, and the release tools.
The natural choice is then to implement \cwASK in a scripting language,
because that is the only way to effectively and productively integrate such a
variety of tools and environments.
The prevalent scripting language in \cwAtlas software is
\cwPython~\cite{ref:python}, which is \cweg\ used for the scripting facilities of
the \cwAthena framework.
\cwPythonCap is often praised for its effective use as a ``glue'' language,
because of its extending and embedding features, and the choice for \cwPython is
therefore natural.

\cwPythonCap is portable, interpreted, and gives you a lot of functionality per
line of code because of its extensive set of ready-to-use modules that come with
every standard installation.
That, and the fact that most (if not all) \cwUnix distributions actually ship
with a version of \cwPython, means that all the user needs to do to install
\cwASK, is to make sure that the scripts are located in a directory that is in
included in their PATH environment variable.
This is the case for all \cwAtlas accounts at CERN, because \cwASK has been
installed in the official scripts directory.
Consequently, all \cwAtlas users have immediate access to \cwASK without the
need of any further setup.

\section{USER INTERFACE}

This section describes how the \cwASK functionality is presented to the user.
The details of the individual commands, buttons \cwetc\ are explained in the
manual~\cite{ref:ASK-manual}, which is included with the distribution, and are
not further elaborated here.

There are two user interfaces in \cwASK: a graphical user interface (GUI) and
a command line interface (CLI).
The former is meant for beginning and casual users, as well as for tutorials;
whereas the latter is more attuned to advanced users, developers, and
librarians.

\subsection{Graphical User Interface}

\begin{figure*}[t]
\centering
\includegraphics[width=0.9\textwidth,angle=0]{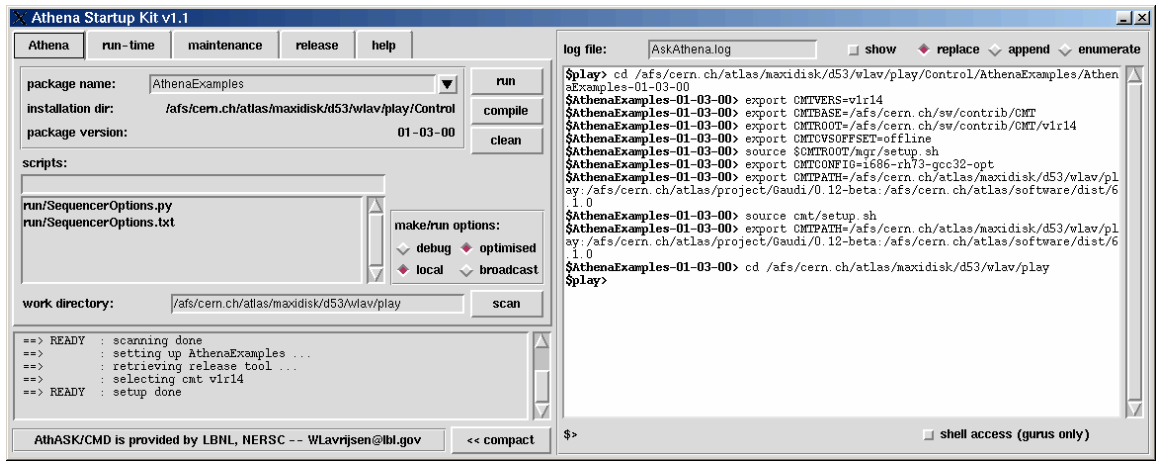}
\caption{The main Athena Startup Kit graphical user interface.}
\label{fig:ASKGUI-full}
\end{figure*}

Historically, the first usage of \cwASK was to generate an \cwAthena skeleton
package.
In order to be able to capture all user-configurable parameters at once, a
small graphical interface interface was developed.
This proved especially practical for package updates, since the current
information could be displayed, ready for the user to modify.
For educational purposes, a pseudo-shell, displaying the underlying commands
that \cwASK executes, was added to the GUI.
A sample view of the full GUI is shown in Figure~\ref{fig:ASKGUI-full}.
On the left, the user can find all tasks, grouped by category in notebook
tabs; on the right is the pseudo-shell, where the user can see the commands that
\cwASK executes and their the results.
In principle, if the user's shell environment is clean, these commands can be
repeated on a real login shell to achieve the same results as with \cwASK.

The categorization of tasks is chosen in such a way as to make sure that a
user only rarely needs to leave the current screen.
In fact, the screen selected on startup is different for an empty work
directory (where the user will most likely want to create or checkout a new
package), than for a directory with existing packages (where the user will
most likely want to work with any of these packages).
There are no dialog boxes in \cwASK: all input, which is minimal and
completely optional by design, is collected directly in the current screen,
and all output messages are in a log window that is always visible.

Integration of interactive \cwAthena is via the pseudo-shell.
The entry box on the lower right automatically activates when the \cwAthena
prompt is detected, and the user can enter input to \cwAthena there.

\subsection{Command Line Interface}

The feature set available through the CLI is the same as the one available
through the GUI.
But there are two main advantages, for proficient users, to the CLI over the
GUI:

\begin{itemize}

\item
The CLI is really just the standard \cwPython interpreter.
Thus, if the user is well versed in \cwPython, she can use any and all
functionality available in \cwPython.
Furthermore, rather than typing all commands, sets of commands can be
collected in a script which is subsequently executed in the interpreter.

\item
Access to underlying system internals, the shell environment, build and
release tools, \cwetc\ is more natural than in the GUI.
One of the options available to the expert is starting a login shell as a
subprocess of \cwASK.
This shell will inherit a properly setup environment, but otherwise acts the
same as if \cwASK wasn't there.

\end{itemize}

An additional advantage, for all users, is that the CLI can be used over a
poor network connection\footnote{With the new version of \cwASK, which can
be installed locally while issuing commands on a secure shell to a remote host,
this is less of an issue.} where the GUI is prohibitively slow.
Finally, the CLI includes a few ``\"{u}ber-recipes,'' \cweg:\\

\noindent {\tt [lxplus]~mkdir~work;~cd~work\newline
\noindent [lxplus]~ask\newline
\noindent ==>~~~~~:~welcome~to~Athena~Startup~Kit~v1.1\newline
\noindent ==>~NOTE:~enter~"help()"~to~receive~help\newline
\noindent >>>~run(~'AtlfastOptions.txt'~)}\newline

\noindent will perform all steps necessary to run the \cwAtlas fast simulation
from the latest official release (for any other package, all the user needs to do
is specify the appropriate, released options file).
It absolutely does not get any easier than this!

Integration with interactive \cwAthena is natural as long as the user realizes that
\cwAthena is run from a separate \cwPython process that does not share any functions
or data with the \cwASK process: the user simply sees a prompt switch and can just
continue typing.

\section{APPLICABILITY}

Naturally, \cwASK is used by the author himself, both to run \cwAtlas software
as well as when working on the development of the \cwAthena framework.
The typical \cwASK user, however, is a beginning user, because of the exposure
of \cwASK to the \cwAtlas collaborators through the software tutorials.
The reconstruction tutorial~\cite{ref:recotutorial} is used by many as their
first starting point with \cwAtlas software: the tutorial shows how to roll your
own \cwAthena algorithm and how to access reconstruction objects in the
transient data store.
That is the typical for how most end-users actually work, thus the
reconstruction tutorial is popular.

\cwASKCap has been in use for some time for the builds of the \cwAtlas software
release at LBNL.
This use exploits the \cwASK error handling in a creative way: when resources
are missing (which they by definition are, since the build has yet to take
place), \cwASK will fall back on CERN, where the release has already been built,
since no build is attempted at LBNL before that.
Thus, \cwASK can fully automatically retrieve the latest settings, versions of
tools, and targets from the CERN release and apply them as appropriate when
building the LBNL release.

\section{OUTLOOK}

It is clear that \cwASK is currently the best way to get started with \cwAtlas
software.
Getting something to run successfully, and being able to look at some output
files with data in them, is a great motivator for users new to \cwAtlas.
When its abundant documentation, its use in tutorials, and its ready
availability are taken into account, then \cwASK's value is obvious.

As \cwAtlas software stabilizes, and as parts of it get replaced by higher
quality software, the number of needed workarounds can be trimmed: at some point
this will render most of \cwASK obsolete.
Elements of it, however, will remain relevant: creating new packages, updating
algorithm and requirements files, its tutorial aspect, \cwetc, all have a long
life expectancy.
It is foreseen, that these parts will be packaged as software components and
transferred to the \cwGanga~\cite{ref:ganga} project, which has already
benefitted from the design ideas from \cwASK.

There will be no major additions of functionality to \cwASK, because of this
expected integration with \cwGanga.
For example, \cwASK does not currently have the capability to submit jobs to
batch systems, even though it is relatively simple to add such a feature based
on the generated standalone execution scripts.
Instead, \cwASK will be maintained to handle future distributions of the
\cwAtlas software and to improve on its existent functionality.
It will keep serving its current audience until it gets replaced by a version
of \cwGanga that has equivalent or better functionality.\\
\vfil

\begin{acknowledgments}
This work was supported by the Office of Science.
High Energy Physics, U.S. Department of Energy under Contract
No.~DE-AC03-76SF00098.
\end{acknowledgments}


\end{document}